# Model-based training of manual procedures in automated production systems

Frieder Loch[1], Gennadiy Koltun[1], Victoria Karaseva[1], Dorothea Pantf¨order[1], Birgit Vogel-Heuser[1]

**Abstract**

Maintenance engineers deal with increasingly complex automated production systems, characterized by increasing computerization or the addition of robots that collaborate with human workers. The effects of changing or replacing components are difficult to assess since there are complex interdependencies between process parameters and components. The introduction of models that describe such dependencies into training systems could support the understanding of these interdependencies and the formation of a correct mental model of a maintenance procedure and the machine and thereby improve the training success. This paper proposes a model-based training system that introduces domain-independent SysML models that formalize such dependencies. The training system consists of a virtual training system for initial training and an online support system for assistance during the procedures. The on- and offline training systems visualize the state of the machine at a certain step of the procedure using structural SysML models. An evaluation of the system against a paper-based manual validated the motivations and showed promising results regarding effectiveness, usability and attractiveness.

## 1. Introduction

Maintenance engineers frequently need to replace components or adjust parameters of industrial machines or robotic cells. This may be caused by the need to exchange defective components or to optimize the manufacturing process. However, each parameter and each component is interconnected with other parameters and components of the machine. The effects of adaptations are therefore hard to determine. Hence, engineers face increasing challenges since the costs of production downtimes caused by delays or errors are high.

The application of virtual reality (VR) and augmented reality (AR) in training and assistance systems is receiving attention from industry and research. Virtual training systems provide cost-efficient and attractive training environments. Augmented reality systems support workers during manual assembly or maintenance procedures [1, 2]. Existing training systems focus on the explanation of work steps and do not visualize the components of a machine and their interdependencies. Models can formalize and visualize the components of automated production systems (aPS) and the interconnections between their parameters. The knowledge that models capture makes the interdependencies between components explicit (e.g. electrical or mechanical dependencies) and can therefore benefit training. The proposed training system uses models based on SysML4*Mechatronics* [3] to describe an aPS and support the training of procedures.

This paper proposes a model-based training system for manual procedures in the industrial domain. The applied models should support the training process by indicating the dependencies between components and providing a structuring mechanism to support the formation of correct mental models. The system should enable the engineer to perform the procedures with less errors and more quickly in the real environment after training. The system consists of a virtual training system for initial training and an online support system that supports the engineer during the work process.

This paper is structured as follows. First, a review of the state of the art of industrial training systems and model-based engineering techniques is provided (see Section 2) to identify the research gaps and derive the requirements for the system (see Section 3). In Section 4 the concept for the training system is proposed and motivated. Section 5 describes a sample scenario that is used in the evaluation in Section 6. A conclusion and future research directions are given in Section 7.

## 2. State of the art

This section provides a review of the state of the of virtual and augmented realitybased training systems. It summarizes their motivations and describe the applied techniques for the visualization of instructions. A discussion motivates the addition of models to training systems as an additional structuring and visualization mechanism. Section 2.2 discusses the use of modeling techniques for aPSs and motivates the application of SysML models.

### *2.1. Industrial training and assistance systems*

Training and assistance systems for workers in industry receive attention from research and industrial practice to address the increasing complexity of industrial machinery. Training systems teach a skill or a procedure. They allow one to get acquainted with a procedure before being trained at a real machine. Assistance systems provide support during the work process and should enable a good performance when carrying out a task for the first time [4]. Assistance systems often target manual assembly procedures and use augmented reality (e.g. [1, 5]).

#### *2.1.1. Virtual training systems*

Virtual training systems simulate industrial procedures and allow practicing tasks without the risks or costs that would be involved when training within the real environment (e.g. using a real machine or interacting with a robotic cell) or with a personal trainer. Virtual training systems should increase the interest and motivation of trainees and, thus, increase the effectivity of the training [6]. Such training system should provide an environment that elicits the feeling of standing in front of the machine and not a screen. This feeling is called presence and can increase the effectivity of a training system since the observed physical responses of the trainee tend towards those that would be observed when interacting with a real machine [7].

A main application of virtual training is when it is not feasible to train with the real tools or within the real context. The training of the assembly processes of a new product can thereby begin before the actual parts have been manufactured [8]. Virtual training of lathing, welding, or CNC-machine operation should mitigate the risk of injuries or of damages to the equipment that a novice may cause [9, 10, 11]. Simulations of spray painting reduce the amount of excessive paint that a novice may spend when using a real spray gun. The system of Konieczny et al. can simulate such processes with high accuracy according to expert evaluations [12]. Borba et al. [13] and Galvan-Bobadilla et al. [14] present systems for the training of potentially dangerous power line maintenance scenarios under different environmental influences.

Training systems for procedures focus on written descriptions of the work steps and visual indications. Galvan-Bobadilla et al. indicate the next work step by a written text and a graphical highlight of the location [14] Ordaz et al. [8] focus on graphical indications of the location where a component needs to be added to the assembly. Brough et al. introduce videos of assembly procedures in a training system [15]. Further training systems introduce further in- and output modalities such as auditory or haptic in- and output. The systems of Guti´errez et al. [16] or Rodr´ıguez et al. [17] propose multimodal approaches that introduce haptic interaction.

*2.1.2. Assistance systems*

Assistance systems are utilized during the work process and should allow novice users to carry out work tasks successfully. They display instructions using augmented reality technologies. Examples are display-based systems that overlay a video stream with graphical visualizations of an assembly procedure [5]. Further systems provide projections of instructions on the work bench [18], or apply head-mounted displays [19].

The research in assistance systems focuses on quantitative comparisons due to the controllable environment within assembly procedures. Such evaluations often target the effectivity of different output technologies. Tang et al. [20] compared augmented reality assistance with text-based manuals and non-registered instructions. Loch et al. [5] compared video-based assembly instructions with display-based augmented reality. Funk et al. [1] compared different output media for AR instructions (e.g. screen or projections). A comparison between an augmented reality system and an electronic manual is provided by Henderson & Feiner [19]. Evaluations typically yield benefits for AR-based assistance systems in comparison to traditional teaching methods like paper-based manuals.

*2.1.3. Discussion*

Existing training systems do not apply models as a visualization technique. The provided indications are focused on the visual or verbal explanation of the next work step. An abstract visualization of the machine and its component is not provided. Models can provide an additional abstract representation and support the training process by providing a framework for the formation of mental models of the trained procedure. Models can furthermore visualize the state of the machine during the procedure (e.g. the connections between the components of a machine).

No combination of a virtual training with an online support system is proposed in the literature. Existing systems exclusively belong to one of both types and fail to leverage synergies. Existing systems are often difficult to adapt to other use cases and domains (e.g.

from procedure to skill training) since they use specific concepts or hardware setups. Table 1 summarizes the state of the art and indicates how the proposed training system addresses the research gaps.

Table 1: Comparison of the characteristics of the proposed training system with the state of the art of industrial training systems (VT = Virtual Training, OS = Online Support).

| | Domain | Models | VT & OS | Adaptability |
|---|---|---|---|---|
| **Virtual training** | | | | |
| Adams et al. [21] | Procedure | - | - | - |
| Antonietti et al. [22] | Skill | - | - | + |
| Bhatti et al. [23] | Procedure | - | - | - |
| Borba et al. [13] | Procedure | - | - | - |
| El-Chaar et al. [24] | Procedure | - | - | 0 |
| Galvan-Bobadilla et al. [14] | Procedure | - | - | + |
| Guti´errez et al. [16] | Procedure | - | - | + |
| Jo et al. [9] | Skill | - | - | - |
| Konieczny et al. [12] | Skill | - | - | - |
| Liang et al. [10] | Skill | - | - | - |
| Loch & Vogel-Heuser [25] | Procedure | - | - | + |
| Ordaz et al. [8] | Procedure | - | - | - |
| Pugmire et al. [26] | Procedure | - | - | 0 |
| Rodr´ıguez et al. [17] | Procedure | - | - | + |
| Wu & Fei [27] | Procedure | - | - | 0 |
| Xiaoling et al. [11] | Skill | - | - | - |
| **Online support** | | | | |
| Aehnelt & Wegner [28] | Procedure | - | - | + |
| Besbes et al. [29] | Procedure | - | - | - |
| Blattgerste et al. [30] | Procedure | - | - | 0 |
| Funk et al. [1] | Procedure | - | - | - |
| Funk et al. [18] | Procedure | - | - | - |
| Henderson & Feiner [19] | Procedure | - | - | + |
| Hoˇrejˇs´ı [31] | Procedure | - | - | - |
| Hou et al. [32] | Procedure | - | - | 0 |
| Korn et al. [33] | Procedure | - | - | - |
| Loch et al. [5] | Procedure | - | - | - |
| Paz et al. [34] | Procedure | - | - | - |
| Tang et al. [20] | Procedure | - | - | - |
| Wang et al. [2] | Procedure | - | - | - |

| This approach | Procedure | + | + | + |
|---|---|---|---|---|

*2.2. Modeling of automated production systems*

A model describes “a simplified version of a concept, phenomenon, relationship, structure or system” [35]. Models provide specific views of a system for different stakeholders from different disciplines (e.g. mechanical engineering or software development). Different formalisms (e.g. graphical, mathematical or physical) are used to represent such an abstraction of the real system.

*2.2.1. Model-driven engineering*

Model-driven engineering (MDE) addresses platform complexity and problems associated with the integration of large-scale systems [36]. Therefore, the role of MDE techniques like UML, a graphical notation for software-based systems, is increasing. The use of profiles allows for the description of the views of different domains on the same object, for instance, the different software aspects of an aPS [37]. Schmidt [38] outlines the benefits of MDE in the software engineering domain compared to third-generation programming languages like JAVA. These benefits, for instance the better expandability and maintainability, can yield a significant decrease of development time. However, MDE techniques are not simply transferable from software engineering to the domain of aPSs. Therefore, the following section introduces approaches for the modeling of an aPS.

*2.2.2. Model-based systems engineering*

The engineering of aPSs is moving from a document-centric procedure towards modelbased systems engineering (MBSE) [39]. MBSE aims at the use of integrated models for “requirements, design, analysis, verification and validation” [40]. Su¨nder et al. [41] applied methods for the verification of modelled system requirements. Secchi et al. [42] proposed a concept for the application of object-oriented and formal models. Bonf`e et al. [43] use models to generate software for aPSs.

A common language in MBSE is Systems Modeling Language (SysML) [44]. SysML is a graphical language that is standardized by the Object Management Group (OMG). It contains structural diagrams (e.g. block definition or internal block diagram) that specify the system’s architecture and behavioral diagrams (e.g. state diagrams) that specify its behavior. Further diagram types allow for the definition of parametric constraints or provide a requirements perspective of the system. SysML can be extended with profiles, for example for the modeling of real-time systems [45] or of specific mechatronic aspects [3, 46, 47]. Besides, SysML models are used to specify the control behavior of aPSs to generate control code automatically [48, 49]. Nevertheless, the understanding of models and their application and benefits in industry is still an open challenge [50, 51] that might be overcome by their application in training systems. To summarize, modeling allows for a focus on specific aspects of a system and the leveraging of synergies between different models if necessary [51].

## 3. Requirements for a model-based training system

The proposed system addresses four requirements: (1) the system applies SysML models in the training process, (2) the system uses a combination of a virtual training system and an online support system, (3) the system targets the training of manual procedures, (4) the system uses no specific hardware and can be adapted to other use cases and domains easily. The requirements are discussed below.

**Application of SysML models.** Current training systems do not apply models as visualizations of the state of the procedure and the machine. The application of models in a training system promises several benefits. It facilitates the understanding of the components and the structure of the machine and thereby supports the formation of a correct mental model. A mental model describes individual assumptions, generalizations and perceptions about reality [52]. The formation of a correct mental model should increase knowledge retention and the effectivity of the training system. It also provides a framework to which users can relate their knowledge to support meaningful learning [53]. Furthermore, the application of a graphical representation allows for the addressing of trainees who prefer a visual learning style. The applied *SysML4Mechatronics* profile was selected since it provides a comprehensible representation of the components of a machine and their interconnections.

Industry mostly discovers drawbacks in applying model-based approaches, although successful research has been presented [50]. These drawbacks include the lack of knowledge and skills to apply MBSE successfully. The authors hypothesize that a model-based training system can emphasize the potentials of MBSE. Through the better understanding of models, the knowledge about aspects such as function-based and systemic thinking might be fostered [54].

**Combination of virtual training and online support.** The existing approaches present either a virtual training system or an online training system. However, a combination of both appears promising. The online support system could reuse content from the virtual training to increase the intuitiveness of the support system or adapt according to the user's performance with the virtual training system. Such an adaptation could provide less online support if the user appears to be experienced during initial training with the virtual training system.

**Adaptability to different use cases.** The training system should be adaptable to the requirements of different use cases and domains (i.e. in terms of effort and cost). Therefore, no specific hardware components should be used. Considering this aspect in the design of a training system facilitates the adaptation of such systems in industrial applications.

## 4. Concept for a model-based training system

This section introduces a concept for a model-based training system. The proposed training system is comprised of a virtual training system and an additional online support system. Both components use SysML models that describe the components of the plant. Section 4.1 introduces the components of the training system. Section 4.2 motivates the didactic approach.

### *4.1. Components of the training system*

The proposed training system combines two components. A virtual training system provides initial training to the users in a virtual environment. The online support system assists the users while carrying out the trained procedure. The following section describes the functions of both components.

#### *4.1.1. Virtual training system*

The virtual training system should provide a realistic and immersive representation of the machine that is targeted by the procedure. Therefore, a three-dimensional model of the machine is the main part of the virtual training system (see Figure 1). The perspective resembles the one that a person would have when standing before the machine and can be freely zoomed and rotated. To increase the immersion of the virtual training system and facilitate the transfer from the virtual environment to the real world [55], the output device is a large and high-resolution display.

A lesson is represented as a linear sequence of work steps. Each step is described by a textual instruction, a graphical annotation, and a structural SysML model. The textual instruction is displayed at the top of the training system. The graphical annotation indicates the location of a step using an arrow that is pointing at the targeted component (see Figure 1).

Models of the current state of the machine are the third part of the instructions. They provide another representation of the machine and support the understanding and the retention of the procedure. The training system uses structural models based on the profile SysML4*Mechatronics* [3] that allow for the interdisciplinary modeling of the mechanical, electric, and software components of an aPS. Such a model describes the components of a machine and their properties (e.g. the resistance of an electric DCmotor or the weight of a mounting plate). Furthermore, the model describes the interfaces between the components and the interdisciplinary dependencies (e.g. whether there are mechanical or electrical connections). The structural model visualizes the state of a machine at different stages of a procedure. The training system highlights connections that have to be removed or established in the given step. Figure 1 shows the main view of the virtual training system.

To allow for the unrestricted movement of the trainee during the training, a tablet application controls the virtual training system via a wireless connection. It allows changing the displayed work step and modifying the perspective of the three-dimensional model on the screen. The user interface elements were designed for eyes-free interaction to allow the trainee to focus on the display of the training system. Since parts of the models may be difficult to read from a distance, the tablet application mirrors the textual instructions and the displayed model if desired. This functionality supports the usability of the system for workers with vision impairments. Figure 2 shows the user interface of the tablet application.

The virtual training system is used in a separated and sheltered environment, for instance, a training center. This reduces distractions that may otherwise interfere with the teaching process. This setting should also provide a suitable training environment for workers with limited capabilities, for instance, aging, low-educated or unskilled workers. Furthermore, the pace of the training system, the complexity of the lessons, and the presentation can be adjusted to the needs of individual trainees [25].

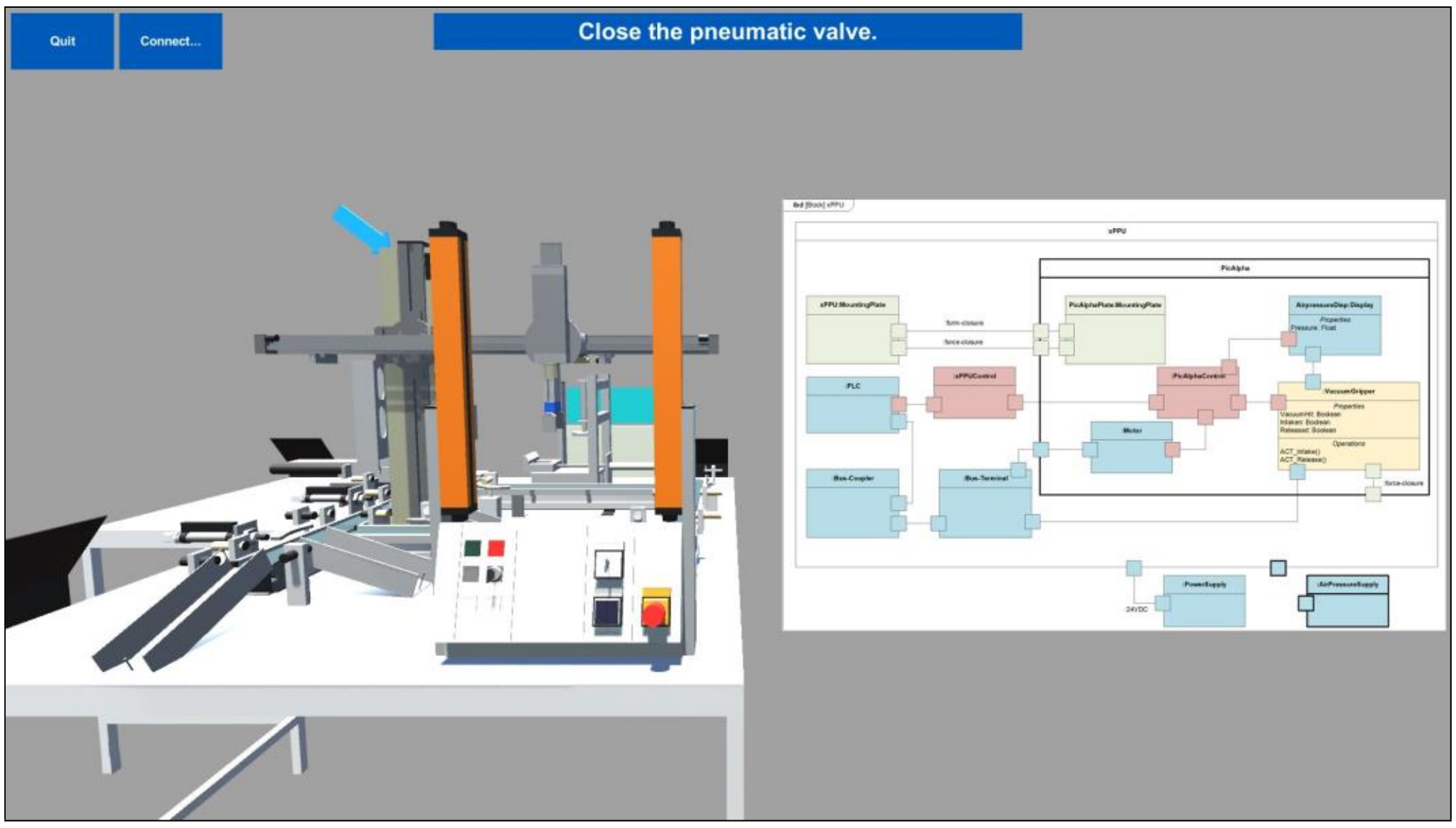

Figure 1: Main view of the virtual training system that shows the three-dimensional plant model and the structural SysML model. The location of a work step is highlighted with an arrow. The SysML model depicts the current configuration of the plant.

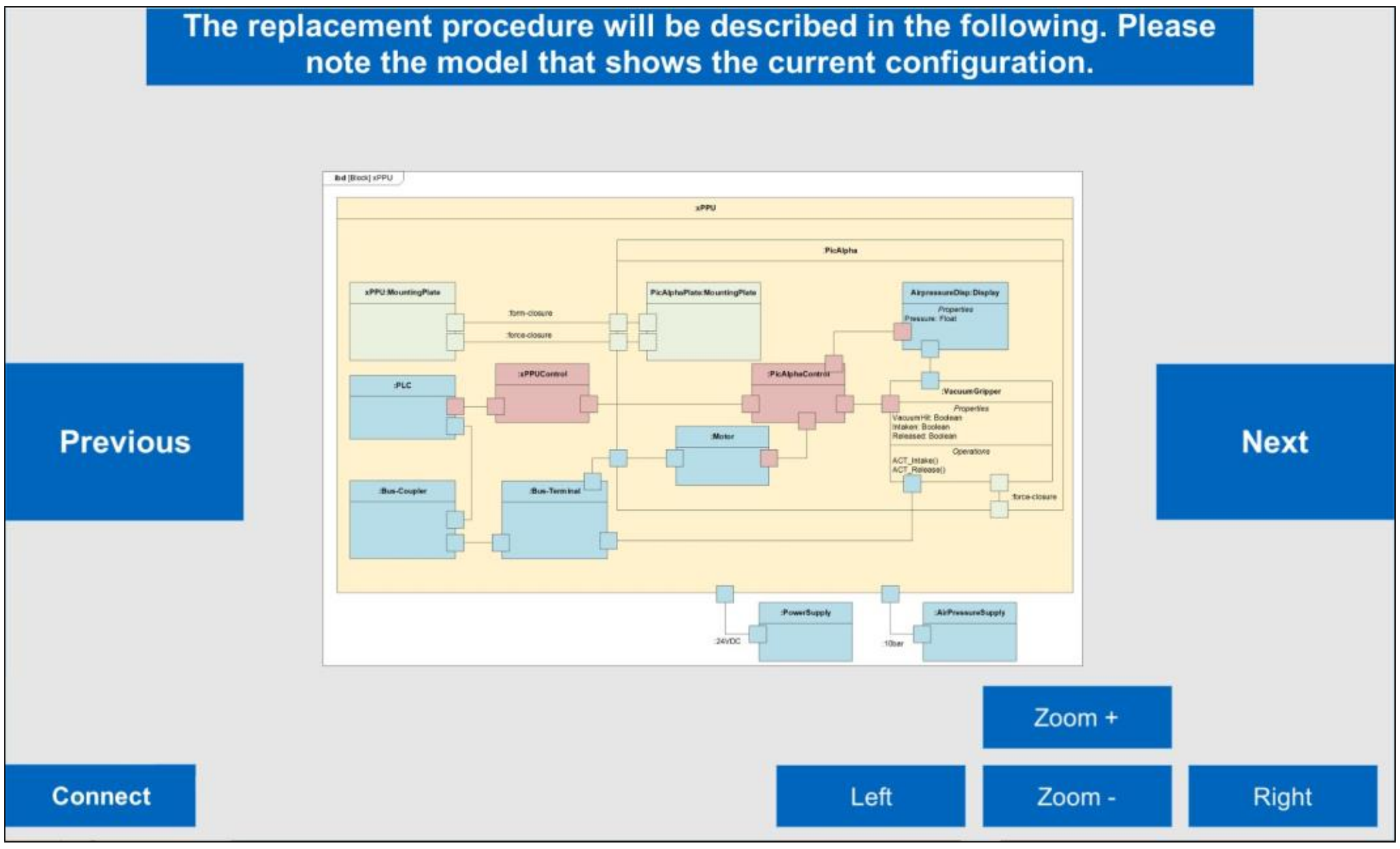

Figure 2: Interface of the tablet application that controls the virtual training system. It mirrors the structural model and allows changing the displayed work step and modifying the perspective of the virtual training system.

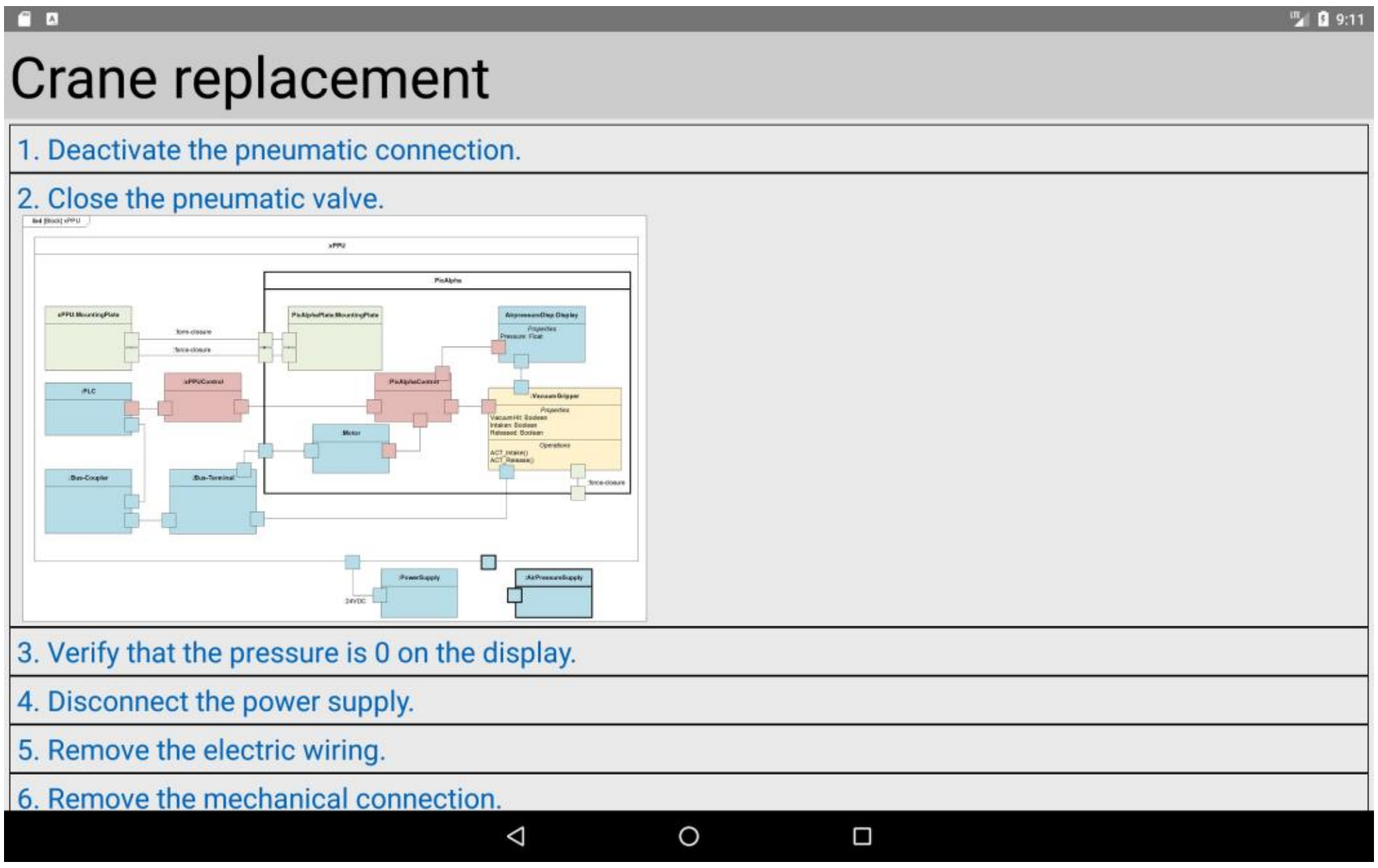

Figure 3: Interface of the online support system. The support system provides a list of instructions for the selected procedure and the structural SysML model of the current work step.

### *4.1.2. Support system*

After having received initial training, a worker may need support when carrying out a procedure in practice. A potential problem that could require referring to a support system could be an uncertainty of either type or location of the next step of the procedure. A support system that can be used if the worker encounters problems during the work process is proposed in addition to the virtual training system.

The support system provides verbal descriptions of the work steps. The structural SysML models can be displayed by the user (see Figure 3). The models from the virtual training systems are used to support the retention of the procedures that were trained with the virtual training system and enhance the intuitiveness of the support system. The user can change the displayed work step.

The user can switch between coarse and detailed instructions. The simplest form displays the list of instructions. At a second level of detail, the structural models are displayed. The support system is provided as an Android-based application for a tablet computer and is operated using the touchscreen.

## *4.2. Motivation of the didactic approach*

The following section motivates the didactic approach of the training system. First, the approach is located within the dichotomy between formal and informal learning. Then the basic teaching strategy is selected by a discussion of common learning theories. The last sections motivate two main features of the presented approach: the use of a virtual environment and the use of structural models within training.

#### 4.2.1. *Formal and informal learning*

There is a prevalent distinction between formal and informal learning. Formal learning is organized and motivated by clear learning objectives [56]. Training systems that teach standardized procedures typically pursue formal learning approaches. Examples are the systems of Ordaz et al. [8] or Gutierrez et al. [16] that teach assembly procedures or the training system for painting of Konieczny et al. [12]. Informal learning is characterized by the absence of a teacher and unstructured or unintentional learning [57]. Participative approaches of informal knowledge-sharing are receiving increasing attention with the transformation of the shop-floor to a knowledge-intensive working place [58] and are typically used for the collection of expert-knowledge, for instance using videos [59].

The proposed training system pursues, in line with the related work, a formal learning approach with explicitly structured lessons. A fixed procedure is taught in a separated learning environment that is optimized for the initial learning of a procedure.

#### 4.2.2. *Learning theories*

Learning theories provide verified instructional strategies for facilitating learning, as well as tools for strategy selection [60]. The selection of a strategy depends on factors such as the level of cognitive processing that is required for the task.

Ertmer and Newby [60] distinguish three basic learning theories: Behaviorism, Cognitivism, and Constructivism. Behaviorism depicts learning as successful if the student shows a certain response when confronted with a stimulus. Training systems that teach standardized procedures (e.g. Funk et al. [1] or Loch et al. [5]) typically apply behavioristic approaches since there is only one permissible way to carry out a procedure and the trainee has to reproduce this sequence exactly.

Cognitivism introduces concepts from cognitive science (e.g. problem solving) and promotes the meaningful storage of information. Introducing models to the training system should leverage the potentials of this strategy. Models that store knowledge about the structural relations and the functional relationships of parts are vital for interaction planning [61]. Possessing or providing such a structuring mechanism is one requirement for the more effective form of meaningful learning which is stressed by cognitivism [53]. Meaningful learning happens when the learned material is conceptually clear and presented in a way that relates it to the prior knowledge of the learner [53]. This can be achieved using formalisms like concept maps [53] or cause-effect graphs [62]. Models should support meaningful learning by providing a framework for the meaningful storage of the trained procedures.

Knowledge about procedures can usually also be applied in other procedures. Different procedures could, for instance, require the removal of an electrical connection. Therefore, training systems supplement their approaches through contextual information that describes the applied tools and related procedures (e.g. El-Chaar et al. [24], Aehnelt & Wegner [28]). The introduction of models in the proposed training system has the aim of providing a framework for the standardization and abstraction of knowledge to facilitate the transfer to similar procedures.

#### 4.2.3. *Virtual environments*

Compared to two-dimensional paper-based training systems, virtual environments provide an increased sense of presence. Presence is defined as the sense of being there in the virtual – instead of in the real – environment. The virtual training system provides several characteristics that enhance presence. Lombard et al. [63] mention aspects like visual display size, image size and quality, dimensionality and the subjective camera perspective. Interaction with a three-dimensional virtual environment increases presence according to Witmer and Singer [64]. Prior studies show an increased ability to detect problems in a process monitoring task if operators were given the possibility to interact with a 3D-representation of process data [65]. The possibility to rotate and zoom the three-dimensional model was, beyond the possibility to view the machine from different angles, meant to increase the engagement of the learner with the application.

Involvement and immersion are facilitated by systems that provide presence. Involvement is the ability to focus on a specific task or a specific subject. Immersion describes the feeling of being physically present in a non-physical environment [66]. Furthermore, a virtual system environment, for instance, desktop PC, VR-Wall or VR-Glasses, improves immersion [67]. The benefit of an immersive system with high presence is that the observed physiological reactions will tend towards those that would be observed in a real environment [7]. In the industrial use case, observing a procedure on a virtual machine would have a similar effect like doing so on a real machine. Presence is reported to have positive effects that could support the learning process. This includes an increase of enjoyment, involvement with the application, memory and skills-training [63].

## 5. Training scenario

After presenting the concept of the model-based training system, a scenario that is used to demonstrate and evaluate the training system is introduced. The scenario is based on a demonstrator of the Institute of Automation and Information Systems of the Technical University of Munich[1], which is introduced in Section 5.1. Section 5.2 discusses the scenario that was chosen for the evaluation. The development of the SysML models that are used in the training system is described in Section 5.3. The technical implementation is summarized in Section 5.4.

### *5.1. The xPPU-demonstrator*

To study the field of evolution in automation, a bench-scale demonstrator was built. This demonstrator handles work pieces of different material and is called the Pick and Place Unit (PPU). The demonstrator consists of a stack, a crane, a stamp and sorting conveyors. Over 18 evolutionary variants and versions of the PPU have been defined. In the last extensions, novel features were installed, such as a large sorting conveyor system, safety cells and a linear handling module, to extend the PPU to the so-called xPPU[2]. Figure 4 shows the model of the demonstrator that was used in the training system.

[1] See https://www.ais.mw.tum.de/en/homepage/.

[2] Further information about the demonstrator is provided by Vogel-Heuser et al. [68].

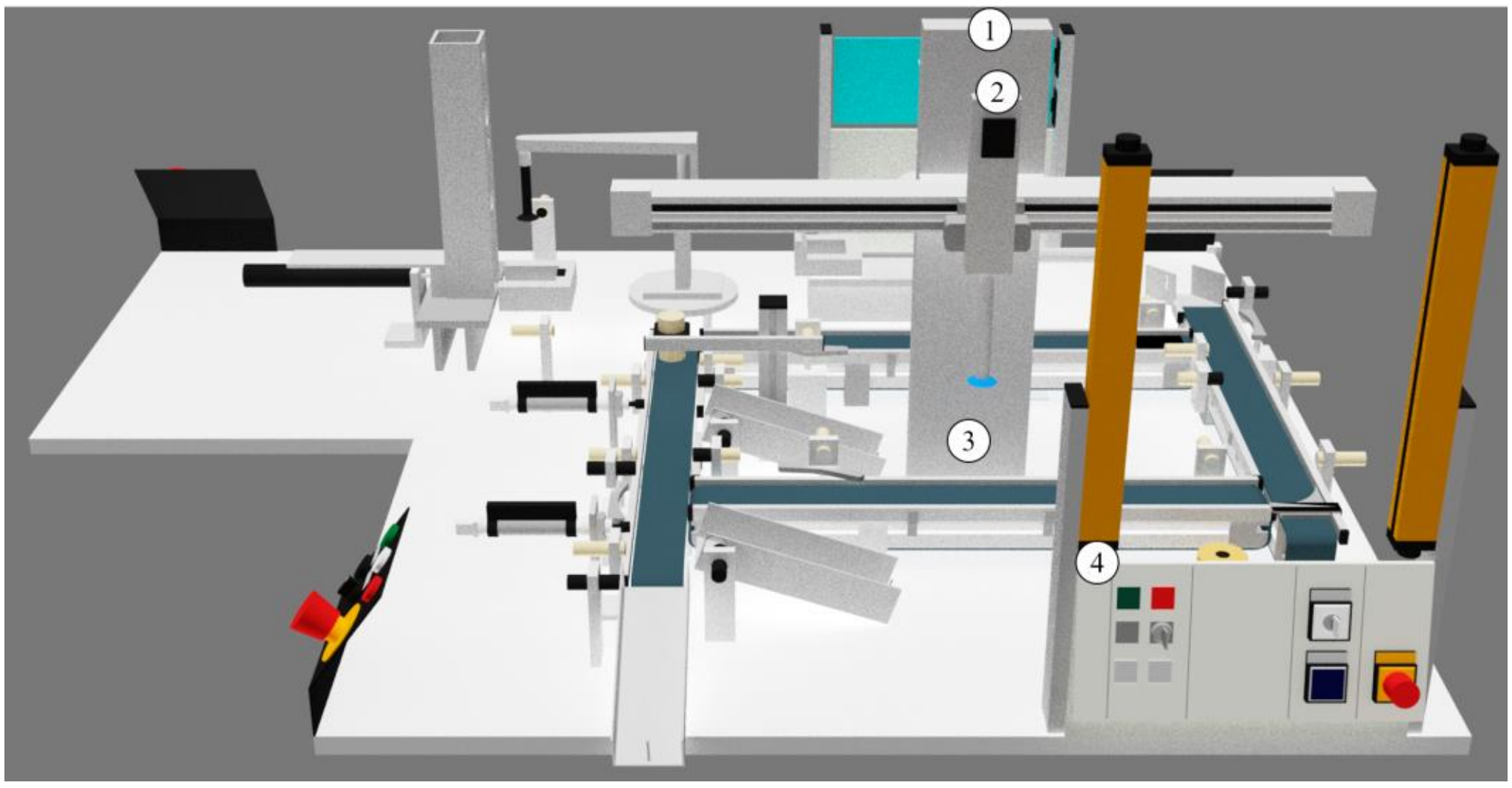

Figure 4: Model of the xPPU-demonstrator that is used in the evaluation of the training system.

### *5.2. Development of a training procedure*

A training procedure that should mimic the complexity of a simple industrial maintenance task was developed. The objective of the training procedure is the replacement of the linear handling module (called *PickAlpha*) by a *Crane* module (see Label 1 in Figure 4). Several motivations to replace a component are possible. The *PickAlpha* module could be malfunctioning or the *Crane* module could be more cost efficient or easier to maintain. Both modules fulfil the requirement to manipulate the processing sequence of work pieces within the sorting conveyor system (see Label 3 in Figure 4). The replacement procedure is described below.

First, the *PickAlpha* module is disconnected from the air pressure supply. This is done in two steps: (1) The supply is disabled in the control software and the respective display is inspected to verify the disconnection (see Label 2 in Figure 4). (2) A valve is closed manually to ensure that no pressured air is provided to the *PickAlpha* module (the valve is located behind the crane base at Label 3 in Figure 4). After that, the procedure continues with the disconnection from the power supply. This means that the operator disconnects all electric connection, which includes unplugging signal communications as well as the power supply cable (see Label 4 in Figure 4). This removes electrical risks for the operator. As the last step, the *PickAlpha* is mechanically disconnected from the xPPU, which is done by releasing the force-closure (remove screws) and the form-closure (remove the module from the mounting plate, see Label 3 in Figure 4).

In the second part, the *Crane* module needs to be installed. This is done in the reversed order. Firstly, the module is mechanically mounted; secondly, the cable connections are closed and, thirdly, the power and air pressure supply is turned on. Additionally, the control software needs to be installed on the PLC and the functionalities of the module and the xPPU need to be tested. This procedure is embedded into the training system and will be used for the evaluation.

The training procedure is transferable to industrial procedures since the correct disconnection and installation of a component is part of the structure of many maintenance procedures. Critical steps of the procedure are the adherence to the sequence of the disconnection of the connections and the inclusion of safety steps (i.e., verifying that the air pressure is 0).

### 5.3. *Modeling of the training procedure*

The following section describes how the structural SysML models of the xPPU were created. The models are based on the SysML4*Mechatronics* profile. A block definition diagram represents the top-level structure and specifies the disciplines to which the components belong according to Kernschmidt & Vogel-Heuser [3]. The internal block diagram (IBD) that represents the xPPU-demonstrator is depicted in Figure 5. A SysML-IBD illustrates (inter-)disciplinary dependencies with ports. These ports outline the affected disciplines of a step of the training procedure.

A simplified IBD of the xPPU and the relevant components for the training procedure is given in Figure 5. This model is focused on the components that are relevant in the given training procedure. This allows for the direction of the attention of the user of the training system to the relevant components. It is hypothesized that the models are, due to the textual labels and the representation of the components and connections, also understandable for users who are not familiar with the language.

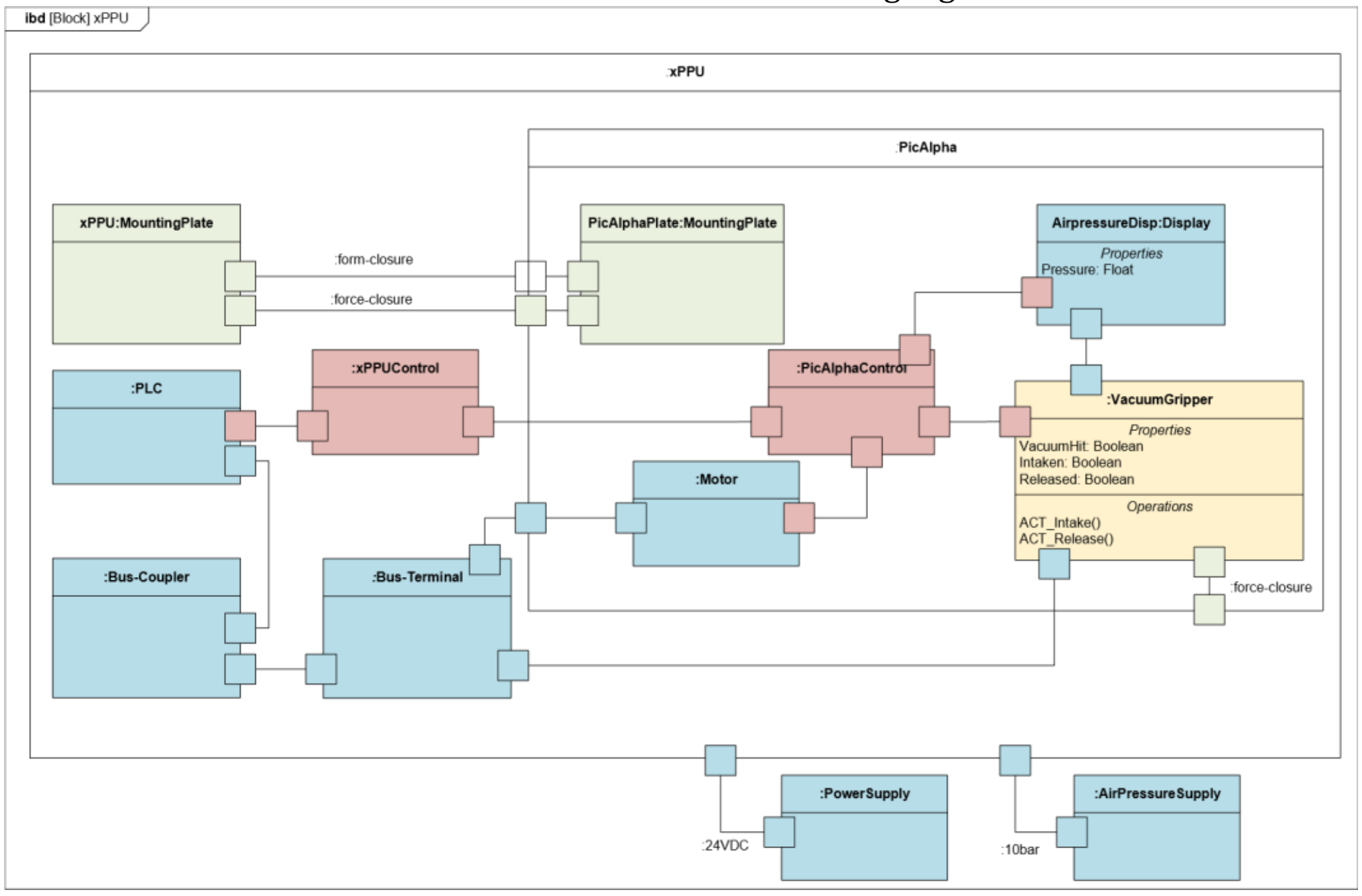

Figure 5: IBD of the xPPU that focuses on the components that are targeted by the training procedure (yellow = mechatronic module, blue = electric/electronic component, red = software component, green = mechanical component).

### *5.4. Implementation of the training system*

The virtual training system and the online support system were implemented using free software to facilitate the implementation and the application in commercial environments. The virtual training system and the remote application were created with the Unity game engine[3]. The model of the demonstrator was created with Blender[4]. The online training system was realized using the Android SDK[5].

The virtual training system is run on a Windows computer that is connected to a large, high-resolution projection screen. The remote application is run on a Microsoft Surface tablet. The communication between both parts should be platform-independent to support the portability to other platforms and the extension of the training system. Therefore, both components exchange plain text messages via a bidirectional socket connection. The online support system is run on an Android-based tablet computer.

## 6. Evaluation of the training system

An evaluation was carried out to gather feedback about the application of SysML models in the training and the online support system. This should provide information for further development, validate the design motivations, and compare the proposed training system with the baseline system of a paper-based manual. The study was set up as a between-subject design. This should allow us to obtain unbiased feedback about the perception of the baseline and the proposed training system and eliminate a possible transfer of skills between the conditions. The participants were split into a group that used the proposed model-based training system and in one group that was trained with a paper-based manual.

The comparison against a paper-based manual was carried out since these manuals are a common training technique in industrial environments. Furthermore, paper-based manuals and virtual training systems can both be used without the involvement of a trainer. Both systems provided the same information and contained the proposed SysML models to obtain broader feedback about their usefulness for different forms of training. Since the variable of interest in this study was the subjective perception of these models, user feedback was used in a qualitative approach to evaluate their comprehensibility and perceived support. The models were assumed to be independent from the context in which they were provided (paper-based or virtual), so the user assessments were generalized across conditions.

### *6.1. Hypotheses*

The evaluation targeted three hypotheses. H1 targets the comparison of the learning performance between a virtual and a paper-based training system. H2 and three subhypotheses target the motivations of the virtual training system (i.e. increased attractiveness, usability, and the elicitation of presence, which is connected to better learning). H3 addresses whether the addition of models improved the training system.

---

[3] https://unity3d.com
[4] https://www.blender.org/
[5] https://developer.android.com/studio/index.html

**H1 – Participants who received virtual training commit less errors than participants who trained using paper-based manuals.** H1 is expected to be valid since the participants who trained with the virtual training system could explore the procedure using an interactive three-dimensional visualization. Especially a better recall of the location of a work step (e.g. the location of the pneumatic connection) was expected due to the three-dimensional representation of the machine.

H1 was validated by measurements of the correct recall of the trained procedure. The influence of individual capabilities was controlled by randomly assigning the participants to the control and the experimental group. H1 indicates whether an approach that is based on a model-based training system promises advantages over a paper-based manual.

**H2 – The proposed training system supports several factors that are assumed to facilitate the learning process: attractiveness, usability, and presence.** The hypotheses is split into three sub-hypotheses that target attractiveness (H2.1), usability (H2.2), and presence (H2.3).

**H2.1 – The proposed training system is more attractive than paper-based training.** H2.1 is expected to be valid since a motivation for the application of VRtechnologies and tablet-based assistance systems is often their attractiveness compared to traditional approaches. H2 was validated by a questionnaire and an interview at the end of the evaluation. The questionnaire was based on the USE-questionnaire of Lund [69] and used the questions that target *Satisfaction*. It is expected that a more attractive system increases the engagement of the trainee and, thus, supports the learning process.

**H2.2 – The proposed training system is more usable than paper-based training.** H2.2 was evaluated by questionnaires. The questions targeting *Ease of Use* from the USE-questionnaire [69] and the *Perceived Ease of Use* questionnaire from Davis [70] were applied. The hypothesis was further supported when the participants did not have to ask for assistance when interacting with the virtual training or the online support system after the introduction. A system with high usability can be used for self-controlled and unsupervised training. It is expected that a usable system facilitates the interaction of the trainee with the system and, therefore, leads to a more effective learning process.

**H2.3 – The virtual training system elicits presence.** H2.3 was validated by a presence questionnaire from Witmer and Singer [64]. Presence is a key motivation for using virtual environments and supports learning (see Section 4.2.3).

**H3 – The introduction of SysML models to virtual and paper-based training systems supports the training process.** H3 was addressed by semi-structured interviews that were conducted after the evaluation. Interviews were expected to yield more valuable insights about the use of models, since specific aspects can be investigated in greater detail. Assessing the benefits and the understandability of the models validates a key motivation of the proposed system.

### *6.2. Participants*

A sample of mechanical engineering students from the second semester was recruited to participate in the evaluation ($n$ = 9, 2 female). The aim of the experiment was to obtain first results about the quality and usability of the proposed training system and the application of SysML models. The considered sample size was considered sufficient for this aim. Such results should motivate industrial partners to support a broader evaluation with participants from the actual target group of industrial workers. The participants were randomly distributed into an experiment and a control group. The experiment group had five members due to a no-show. Participants received a compensation of 15e.

### 6.3. Experimental steps

In the beginning, the participants were briefed about the aim and the process of the evaluation. Members of the experiment group were told that they would test the effectivity of a novel training system. Members of the control group were told that they would test the memorability of paper-based instructions.

After filling in a demographic questionnaire, both groups conducted the training. They were asked to train and remember a procedure of thirteen steps (e.g. "Deactivate the pneumatic connection."). They were instructed that they would have to go through the procedure after the training at the demonstrator and explain the type and the location of the steps as accurately as possible. A standardized verbal introduction and demonstration of the virtual training system was provided before the training. It was expected that this was sufficient due to the simplicity of the interface of the training system. Observations during the training confirmed that the participants successfully used all functions of the system. The participants could use the respective training system for at most five minutes. A time limit was imposed to ensure the comparability of the results of the evaluation of both systems. Five minutes proved to be sufficient in the pilot test, as well as in the evaluations. The participants were allowed to end the training if they felt they were sufficiently acquainted with the procedure, which three out of the nine participants did.

The participants were asked to recall the procedure on the xPPU demonstrator. This was done by having the participant verbally explain the steps that are necessary to carry out the task and indicate the location. It was recorded whether a participant remembered the location (Where?) and the type (What?) of the steps. The participants were told that they may consult the support system in case of troubles but doing so would be noted as a "penalty".

After finishing the recall task, the participants were asked to fill in questionnaires that targeted the attractiveness of the system. Members of the experiment group were asked to fill in an additional questionnaire that measured presence.

Feedback regarding the interface of the system and ideas for improvement were gathered at the end, to obtain additional insights and inform further development. Therefore, the evaluation was concluded by a semi-structured interview. Within the interview, the participants were asked to name positive or negative aspects of the training system and whether they understood the SysML models and whether they supported them in learning the procedure. Suggestions for improvements of the training system were collected as well.

### 6.4. Discussion and results

The results of the evaluations confirm the motivations of the training system regarding performance, attractiveness, usability and presence. Furthermore, the application of models was perceived positively. The conclusions should be validated with a group of industrial users, to test whether they have a similar perception of the training system. The participants did not refer to the online training system. This was possibly due to the evaluation procedure that was too short and also since there was no reason to do so in case of doubts about the correct procedure. An evaluation with a more complex procedure is necessary to target this aspect of the proposed system in detail. The following section discusses the results regarding the three hypotheses.

**H1 – Participants who received virtual training commit less errors than participants who trained using paper-based manuals.** The results show a lower number of errors for the participants of the experimental group (0.8 with the virtual training system, 2.6 with the paper-based system). However, due to the small sample size this can only be taken as an indication of superior performance.

The motivations of model-based training were supported by the comments that the participants made in the interviews. They claimed that the model-based training system allows for a clearer description of the locations of actions compared to verbal explanations or the paper-based system.

**H2.1 – The virtual training system is more attractive than paper-based training.** The results of the questionnaires support this hypothesis. The results of the satisfaction questionnaire (see Table 2) indicate a more positive perception of the paper-based training system in total with a median of 5 for the virtual training system, compared to a median of 3 for the paper-based training. The good scores regarding presence (see H4) may also support the attractiveness of the system. Summarizing, the results of the evaluation seem to validate this hypothesis.

Table 2: Results of the satisfaction questionnaire (Paper-based manual: *n* = 4; Virtual training system (VTS): *n* = 5), (1 = Strongly disagree, 6 = Strongly agree)

| | Paper | | | VTS | | |
|---|---|---|---|---|---|---|
| | Mean | Median | SD | Mean | Median | SD |
| I am satisfied with it. | 3 | 3.25 | 1.4 | 5 | 4.6 | 0.5 |
| I would recommend it to a friend. | 3 | 3 | 0.9 | 5 | 5 | 0.9 |
| It is fun to use. | 2 | 2 | 0.9 | 5 | 5.2 | 0.4 |
| It works the way I want it to work. | 3 | 3 | 0.8 | 5 | 5 | 0.6 |
| It is wonderful. | 3 | 2.75 | 0.5 | 5 | 4.8 | 0.4 |
| I feel I need to have it. | 2.5 | 2.5 | 0.8 | 4 | 4.6 | 0.8 |
| It is pleasant to use. | 2 | 2.25 | 0.9 | 5 | 5 | 0.6 |
| | 3 | 2.64 | 1.07 | 5 | 5 | 0.67 |

**H2.2 – The virtual training system is more usable than paper-based training.** The virtual training system appeared to be intuitive. The participants did not report problems with using the training system. Furthermore, the questionnaires (see Table 3 and Table 4) indicate a good usability of the system (e.g. the items that target ease of use or understandability). Also, the comparison with the paper-based manual yields that the virtual training system was perceived superior. The outcomes of the evaluations support this hypothesis.

There are distinct aspects in which the virtual training system is superior. While receiving similar scores in items for simplicity or ease of use, the virtual training system showed benefits in flexibility, the applicability for first time users, and the understandability of instructions. These benefits may serve as guidelines for the further development and enhancement of the training system.

Table 3: Results of the *Perceived Ease of Use* questionnaire (Paper-based manual: $n$ = 4; Virtual training system: $n$ = 5), (1 = Strongly disagree, 6 = Strongly agree)

| | Paper-based manual | | | Virtual training system | | |
|---|---|---|---|---|---|---|
| | Mean | Median | SD | Mean | Median | SD |
| It is confusing. | 3.5 | 3 | 1.2 | 2 | 1.6 | 0.5 |
| It is error prone. | 4 | 3.5 | 0.9 | 2 | 1.6 | 0.5 |
| It is frustrating. | 4 | 3.25 | 1.3 | 1 | 1 | 0 |
| I need the manual often when using the system. | 5 | 5 | 1 | 1 | 1.4 | 0.5 |
| Using it costs a lot of mental effort. | 5 | 4.5 | 0.9 | 1 | 1.8 | 1.2 |
| I find it easy to recover from errors. | 2.5 | 2.75 | 0.8 | 4 | 4.4 | 0.5 |
| It is rigid and inflexible. | 4 | 4 | 1.4 | 2 | 1.8 | 0.7 |
| It is controllable. | 5 | 5 | 0 | 6 | 5.4 | 0.8 |
| It shows uncontrollable behavior. | 4 | 4 | 1.4 | 2 | 2 | 1.1 |
| It is cumbersome. | 2 | 2.25 | 1.1 | 1 | 1.6 | 0.8 |
| It is understandable. | 5 | 4.75 | 1.1 | 6 | 5.6 | 0.8 |
| It is easy to remember. | 3 | 3.25 | 1.3 | 5 | 5.4 | 0.5 |
| It provides guidance. | 4 | 4.5 | 0.9 | 5 | 4.8 | 0.4 |
| It is easy to use. | 4 | 4.25 | 1.7 | 6 | 5.6 | 0.5 |
| | 3.9 | 3.9 | 1.4 | 3.1 | 3.1 | 0.75 |

**H2.3 – The virtual training system elicits presence.** The results of the presence questionnaire indicate that the virtual training system was able to elicit a sense of presence. This was also further supported by the answers in the concluding interviews. Participants stated that the size of the screen and the perspective by which it showed the model made them feel like “standing in front of the machine”. Hence, the transfer of the procedure from the virtual to the real environment was facilitated. These results appear promising given the fact that the control mechanism (i.e., buttons on a tablet application) and the fidelity of the model were simple. The simplicity of the control mechanisms is also visible in the questionnaire, where the participants indicated that they were aware of the display and control devices (median of 5). The results of the evaluation appear to support the hypothesis and the design motivations of the virtual training system.

**H3 – The introduction of models to virtual and paper-based training systems supports the training process.** The results of the interviews support this hypothesis. The participants (only one participant reported modeling experience) reported that they understood the purpose and meaning of the SysML models during the post-test interviews. The participants expressed that the models provided another mechanism to represent functional and structural connections and dependencies between the components of the plant. Trainees were able to locate a component by its function (e.g. the pneumatic valve of the crane) and not only by its location. This provided additional support, especially for participants with technical experience. Participants also expressed that the models implied a more meaningful structuring of the work steps into functional units (e.g. by grouping the steps that are necessary for removing the pneumatic connection). Overall, the feedback that was obtained in the interviews suggests that these models and the structure they provide can be a facilitating factor in knowledge acquisi-tion, which again suggests that these qualitative effects should be investigated further; for example, through an evaluation with a standardized rating system or a quantitative comparison of how the presence or absence of models affects the learning process in terms of performance.

Table 4: Results of the *Ease of Use* questionnaire (Paper-based manual: *n* = 4; Virtual training system: *n* = 5), (1 = Strongly disagree, 6 = Strongly agree)

| | Paper-based manual | | | Virtual training system | | |
|---|---|---|---|---|---|---|
| | Mean | Median | SD | Mean | Median | SD |
| It is easy to use. | 5.5 | 5.5 | 0.5 | 6 | 5.6 | 0.5 |
| It is simple to use. | 5.5 | 5.5 | 0.5 | 5 | 4.5 | 0.5 |
| It is user friendly. | 4.5 | 4 | 1.2 | 6 | 5.4 | 0.8 |
| It requires the fewest steps possible to accomplish what I want to do with it. | 5 | 4.5 | 1.5 | 5 | 4.8 | 0.7 |
| It is flexible. | 2.5 | 3.25 | 1.6 | 5 | 4.6 | 0.5 |
| Using it is effortless. | 2.5 | 3.25 | 1.6 | 5 | 4.6 | 0.5 |
| I can use it without written instructions. | 3 | 3 | 1.6 | 5 | 5.2 | 0.7 |
| I don't notice any inconsistencies as I use it. | 4.5 | 4.25 | 0.8 | 5 | 5 | 0.9 |
| Both occasional and regular users would like it. | 3.5 | 3.75 | 0.8 | 5 | 4.8 | 0.4 |
| I can recover from mistakes quickly and easily. | 3.5 | 3.75 | 0.8 | 5 | 5.4 | 0.5 |
| I can use it successfully every time. | 4.5 | 4.5 | 0.5 | 5 | 5 | 0 |
| | 4 | 4.11 | 1.4 | 5 | 5.11 | 0.68 |

Table 5: Results of the *Presence* questionnaire for the virtual training system (*n* = 5), (1 = not at all, 6 = a lot)

| | Mean | Median | SD |
|---|---|---|---|
| How much were you able to control events? | 5.6 | | |
| How responsive was the environment to actions that you performed? | 5.6 | | |
| How natural did your interaction with the environment seem? | 5.2 | | |
| How natural was the mechanism that controlled movement through the environment? | 4.4 | | |
| How aware were you of events occurring in the real world around you? | 4.2 | | |
| How aware were you of your display and control devices? | 5.2 | | |
| Were you able to anticipate what would happen next in response to the actions that you performed? | 5.6 | | |
| How completely were you able to actively survey or search the environment using vision? | 4.8 | 5 | 0.4 |
| How compelling was your sense of moving around inside the virtual environment? | 4.6 | 5 | 0.5 |
| How closely were you able to examine objects? | 4.2 | | |
| How well could you examine objects from multiple viewpoints? | 4.4 | | |
| To what degree did you feel confused or disoriented at the beginning of breaks or at the end of the experimental session? | 2 | | |
| How involved were you in the virtual environment experience? | 4.8 | | |
| How distracting was the control mechanism? | 2.8 | | |
| How much delay did you experience delays between your actions and expected outcomes? | 2.2 | | |
| How quickly did you adjust to the virtual environment experience? | 5 | | |
| How proficient in moving and interacting with the virtual environment did you feel at the end of the experience? | 5.4 | | |
| How much did the visual display quality interfere or distract you from performing assigned tasks or required activities? | 1.8 | 1 | 1.2 |
| How much did the control devices interfere with the performance of assigned tasks or with other activities? | 3 | 3 | 0.9 |
| How well could you concentrate on the assigned tasks or required activities rather than on the mechanisms used to perform those tasks or activities? | 5.4 | 5 | 0.5 |
| | 4.6 | 4.3 | 0.6 |

## 7. Conclusion and future work

This paper presented a model-based training system consisting of a virtual training system and an online support system. Structural SysML models were introduced to support the understanding of the interdependencies between the components of the machine and thereby support the formation of a correct mental model of the procedure. The addition of models was perceived positively in an evaluation. The results indicate that the virtual training system improved the learning process and was perceived to be more attractive and usable than the paper-based manual that served as a baseline system. The benefits of the models as an additional structuring mechanism and information source that motivated their inclusion into the training system were supported by the evaluation. An additional evaluation with industrial users is planned in the future. The combination of an virtual training system with an online support system is especially valuable in applications where the costs of errors are high compared to the time for the consultation of a support system. The proposed training system is not limited to the applied modeling language and transferable to other domains. Section 7.1 discusses the application of model-based training in the domain of human-robot-collaboration, where the training system can simulate and visualize interactive collaboration processes. Section 7.2 presents further extensions of the training system.

### *7.1. Applications in human-robot collaboration*

Several requirements for the acceptance of human-robot collaboration in industry have been identified. One factor is a safe interaction between a worker and the plant. Different active and passive systems to avoid collisions or mitigate their results have been proposed [71]. The training and empowerment of operators is another factor and key enabler for safe human-robot collaboration and its acceptance in the industry [72].

Virtual training systems can play an important role in the training of human-robot collaboration. Virtual environments allow for realistic simulation of a robot in a training system for the training of standard and emergency procedures. This allows for the initial training of procedures that involve collaborations with a robot in a secure environment. Thereby, the users are granted time to become familiar with their non-human team members and learn about their interaction style.

As indicated by the results of the reported evaluation, models can be a useful addition for the development of training systems for the interactions between humans and robots. Especially the communication of the intentions of the robot and its expectations about the next action of the human is crucial [73]. Further modeling and visualization techniques that increase transparency could therefore be included in the training system. UML state charts, for instance, can indicate the action that the robot expects from the human or the next actions that the robot will carry out. Similarly, an activity diagram could provide an overview of the state of an interaction between human and robot. Introducing such mechanisms into a training system may enhance the understanding of robot behavior and thereby increase the acceptance of HRC.

### *7.2. Extensions of the training system*

Several extensions are planned to provide a more present and immediate interaction. One approach is the introduction of motion tracking. Changing the perspective according to the position of the user could allow one to experience the trained procedures with more realism. Gesture-based interaction would remove the necessity of a control device for the training system. VR-glasses could improve the training with large machines, since the user could walk around the virtual machine and interact with its different parts. Introducing stereoscopic display could make the training system more effective as well [7]. Since presence benefits from the inclusions of additional media [63], the inclusion of sound to simulate the noises of the factory environment could be beneficial too.

People differ in knowledge acquisition techniques and mindset. Such differences, for instance, due to gender [74], could be accommodated by a training system. Tailoring a training system to individual assumptions about a machine or process promises a more effective training. Such assumptions can be described by mental models [52]. The collection and analysis of the mental models of trainees and experienced professionals is promising since the identification of inconsistencies and contradictions between these mental models could guide the training process [75].

## Acknowledgments

This research was developed within the INCLUSIVE project, founded from the European Communitys Horizon 2020 Research and Innovation Programme (H2020-FoF-042016) under grant agreement n. 723373.

This work was supported by the German Research Foundation (Deutsche Forschungsgemeinschaft, DFG) as a part of the collaborative research centre Sonderforschungsbereich SFB768 “Managing cycles in innovation processes–Integrated development of product-service-systems based on technical products”.